\documentstyle[prl,aps]{revtex}
\twocolumn

\begin{document}
\author{Jian Qi Shen $^{1,2}$ \footnote{E-mail address: jqshen@coer.zju.edu.cn}}
\address{$^{1}$  Centre for Optical
and Electromagnetic Research, Joint Research Centre of Photonics
of the Royal Institute of Technology (Sweden) and Zhejiang
University, Zhejiang University,
Hangzhou Yuquan 310027, P.R. China\\
$^{2}$ Zhejiang Institute of Modern Physics and Department of
Physics, Zhejiang University, Hangzhou 310027, P.R. China}
\date{\today}
\title{A note: the relativistic transformations for the optical constants of
media}
 \maketitle

\begin{abstract}
The Lorentz transformations for the optical constants (electric
permittivity, magnetic permeability and index of refraction) of
moving media are considered.
\end{abstract}

{\bf 1. Introduction}

Some novel vacuum effects associated with the dynamical quantities
(such as energy, spin and momentum) of zero-point fluctuation
fields have been considered in the literature. These effects
include the Casimir effect (related to the vacuum energy and mode
distribution structure)\cite{Casimir,Lam}, magnetoelectric
birefringences of the quantum vacuum\cite{Rikken}, vacuum-induced
Berry's phase of spinning particles (related to the quantum vacuum
fluctuation)\cite{Fuentes} and quantum-vacuum geometric phase of
zero-point field in a coiled fiber system (related to the spin of
vacuum)\cite{Shenpla,Shen}.

As to the effect associated with the momentum of vacuum zero-point
fields, more recently, Feigel has considered the quantum vacuum
contribution to the momentum of electromagnetic
media\cite{Feigel}. We think that the relativistic transformations
for the optical constants (electric permittivity and magnetic
permeability) of the medium should be taken into account when
writing the Lagrangian of the moving electromagnetic system. It
can be shown that the effect arising from such a transformation
will also provide a quantum vacuum contribution to the velocity of
media, in addition to the one derived by Feigel
himself\cite{Feigel}. This Note presents the relativistic
transformations for the optical constants of moving media.
\\ \\
{\bf 2. The relativistic transformations for magnetic permeability
and electric permittivity}
\\ \\
{\it The transformations of ${\bf E}$, ${\bf B}$, ${\bf H}$ and
${\bf D}$} Assume that an inertial frame of reference moves at
velocity ${\bf v}$ relative to a rest system in which an
electromagnetic medium (homogeneous and isotropic) is fixed. In
the following, all the physical quantities with a prime refer to
those in the moving frame while the ones without a prime refer to
the quantities in the rest frame. According to SR, the
relativistic transformation for the electric field strength ${\bf
E}$ and the magnetic induction ${\bf B}$ is of the form
\begin{equation}
\left\{
\begin{array}{ll}
&  {\bf E}'=\gamma \left({\bf E}+{\bf v}\times{\bf B}\right),      \\
&   {\bf B}'=\gamma \left({\bf B}-\frac{1}{c^{2}}{\bf v}\times{\bf
E}\right),
\end{array}
\right. \label{set1}
\end{equation}
and the relativistic transformation for the magnetic field
strength ${\bf H}$ and the electric displacement vector ${\bf D}$
takes the following form
\begin{equation}
\left\{
\begin{array}{ll}
&  {\bf H}'=\gamma \left({\bf H}-{\bf v}\times{\bf D}\right),  \\
&   {\bf D}'=\gamma \left({\bf D}+\frac{1}{c^{2}}{\bf v}\times{\bf
H}\right),
\end{array}
\right. \label{set2}
\end{equation}
where $\gamma$ denotes the relativistic factor. These two sets of
equations will be employed to derive the transformation
relationships of the optical constants between different frames of
reference.
\\ \\
{\it The transformation of magnetic permeability} According to the
relation ${\bf H}'=\gamma \left({\bf H}-{\bf v}\times{\bf
D}\right)$, one can obtain
\begin{equation}
\mu'{\bf H}'=\gamma \mu'\left({\bf H}-\epsilon{\bf v}\times{\bf
E}\right).   \label{mmuu}
\end{equation}
In the meanwhile, the relation ${\bf B}'=\gamma \left({\bf
B}-\frac{1}{c^{2}}{\bf v}\times{\bf E}\right)$ can be rewritten as
\begin{equation}
\mu'{\bf H}'=\gamma \left(\mu{\bf H}-\frac{1}{c^{2}}{\bf
v}\times{\bf E}\right).   \label{eqmmuu}
\end{equation}
Thus, it follows from Eqs. (\ref{mmuu}) and (\ref{eqmmuu}) that
\begin{equation}
\mu'\left({\bf H}-\epsilon{\bf v}\times{\bf E}\right)=\mu{\bf
H}-\frac{1}{c^{2}}{\bf v}\times{\bf E}.     \label{vectorrelation}
\end{equation}
Here, for convenience and simplicity, we consider a time-harmonic
electromagnetic wave propagating along the relative velocity ${\bf
v}$ between the two inertial frames. The wave vector of such a
wave is pointed opposite to the direction of ${\bf v}$. So, the
set of vectors (${\bf v}, {\bf H}, {\bf E}$) form a right-handed
system. In this sense, the direction of the term $-\epsilon{\bf
v}\times{\bf E}$ in (\ref{vectorrelation}) is parallel to that of
${\bf H}$. If $v, H, E$ are the magnitudes of ${\bf v}, {\bf H},
{\bf E}$, respectively, then the magnitude of the term
$-\epsilon{\bf v}\times{\bf E}$ is $\epsilon vE$. Thus we have the
relation
\begin{equation}
\mu'\left(H+\epsilon{v}{E}\right)=\mu{H}+\frac{1}{c^{2}}{v}{E},
\end{equation}
and consequently obtain
\begin{equation}
\mu'=\frac{\mu{H}+\frac{1}{c^{2}}{v}{E}}{H+\epsilon{v}{E}}.
\label{consequently}
\end{equation}
By substituting the relation $E=\sqrt{\frac{\mu}{\epsilon}}H$ for
the time-harmonic wave into Eq. (\ref{consequently}), one can
obtain the relativistic transformation of the magnetic
permeability, {\it i.e.},
\begin{equation}
\mu'=\sqrt{\frac{\mu}{\epsilon}}\left(\frac{\sqrt{\epsilon\mu}+\frac{v}{c^{2}}}{1+\sqrt{\epsilon\mu}v}\right),
\label{mu}
\end{equation}
or
\begin{equation}
\mu'_{\rm r}=\sqrt{\frac{\mu_{\rm r}}{\epsilon_{\rm
r}}}\left(\frac{\sqrt{\epsilon_{\rm r}\mu_{\rm
r}}+\frac{v}{c}}{1+\sqrt{\epsilon_{\rm r}\mu_{\rm
r}}\frac{v}{c}}\right)=\sqrt{\frac{\mu_{\rm r}}{\epsilon_{\rm
r}}}\tanh\left(\sqrt{\epsilon_{\rm r}\mu_{\rm
r}}+\frac{v}{c}\right),    \label{eq9}
\end{equation}
where $\mu'_{\rm r}$, $\mu_{\rm r}$ and $\epsilon_{\rm r}$ are,
respectively, the relative magnetic permeabilities and
permittivity, which are so defined that $\mu'_{\rm
r}=\frac{\mu'}{\mu_{0}}$, $\mu_{\rm r}=\frac{\mu}{\mu_{0}}$ and
$\epsilon_{\rm r}=\frac{\epsilon}{\epsilon_{0}}$.
\\ \\
{\it The transformation of electric permittivity}  The relation
${\bf D}'=\gamma \left({\bf D}+\frac{1}{c^{2}}{\bf v}\times{\bf
H}\right)$ can be rewritten as
\begin{equation}
\epsilon'{\bf E}'=\gamma\left(\epsilon{\bf E}+\frac{1}{c^{2}}{\bf
v}\times{\bf H}\right).
\end{equation}
According to the relation ${\bf E}'=\gamma \left({\bf E}+{\bf
v}\times{\bf B}\right)$, one has
\begin{equation}
\epsilon'{\bf E}'=\gamma \epsilon'\left({\bf E}+\mu{\bf
v}\times{\bf H}\right).
\end{equation}
Thus the following relation is derived
\begin{equation}
\epsilon'\left({\bf E}+\mu{\bf v}\times{\bf H}\right)=\epsilon{\bf
E}+\frac{1}{c^{2}}{\bf v}\times{\bf H},
\end{equation}
which can be rewritten in the form
\begin{equation}
\epsilon'\left({E}+\mu{v}{H}\right)=\epsilon {E}+\frac{1}{c^{2}}{
v}{H}
\end{equation}
under the condition that the set of vectors (${\bf v}, {\bf H},
{\bf E}$) form a right-handed system. So, the permittivity
observed in the moving frame is
\begin{equation}
\epsilon'=\frac{\epsilon
{E}+\frac{1}{c^{2}}{v}{H}}{{E}+\mu{v}{H}}.
\label{movingpermittivity}
\end{equation}
Insertion of the relation $E=\sqrt{\frac{\mu}{\epsilon}}H$ into
Eq. (\ref{movingpermittivity}) yields
\begin{equation}
\epsilon'=\sqrt{\frac{\epsilon}{\mu}}\left(\frac{\sqrt{\epsilon\mu}+\frac{v}{c^{2}}}{1+\sqrt{\epsilon\mu}v}\right),
\label{epsilon}
\end{equation}
or
\begin{equation}
\epsilon'_{\rm r}=\sqrt{\frac{\epsilon_{\rm r}}{\mu_{\rm
r}}}\left(\frac{\sqrt{\epsilon_{\rm r}\mu_{\rm
r}}+\frac{v}{c}}{1+\sqrt{\epsilon_{\rm r}\mu_{\rm
r}}\frac{v}{c}}\right)=\sqrt{\frac{\epsilon_{\rm r}}{\mu_{\rm
r}}}\tanh\left(\sqrt{\epsilon_{\rm r}\mu_{\rm
r}}+\frac{v}{c}\right),    \label{eq16}
\end{equation}
where $\epsilon'_{\rm r}$ ($\equiv\frac{\epsilon'}{\epsilon_{0}}$)
denotes the relative permittivity of the medium observed in the
moving frame.
\\ \\

It should be noted that here the relativistic transformations for
the permeability and permittivity refer only to those along the
directions perpendicular to the wave vector (or ${\bf v}$). The
components of the permeability and permittivity along the wave
vector (or ${\bf v}$) does not alter under the Lorentz
transformation.
\\ \\
{\bf 3. Discussions}
\\ \\
{\it The trivial cases} It follows from Eqs. (\ref{eq9}) and
(\ref{eq16}) that when the relative velocity $v$ between the two
frames vanishes, the permeability and permittivity
\begin{equation}
\mu'_{\rm r}=\mu_{\rm r},   \quad   \epsilon'_{\rm
r}=\epsilon_{\rm r}.
\end{equation}
It is a trivial case deserving no consideration. In addition, for
a vacuum medium in which the relative permeability and
permittivity equal the unity, {\it i.e.}, $\mu_{\rm
r}=\epsilon_{\rm r}=1$, Eqs. (\ref{eq9}) and (\ref{eq16}) give
\begin{equation}
\mu'_{\rm r}=1,   \quad   \epsilon'_{\rm r}=1.
\end{equation}
This means that both the permeability and the permittivity of the
vacuum medium are invariant under the Lorentz transformation
(which is in connection with the principle of invariance of light
speed). In this sense, vacuum can be considered a
Lorentz-invariance medium, all the optical constants of which
measured by different observers who move at relative velocities
with respect to each other are correspondingly the same.
\\ \\
{\it Optical refractive index} The optical refractive index of the
medium observed from the moving frame and the rest frame are
expressed by $n'=\sqrt{\epsilon'_{\rm r}\mu'_{\rm r}}$ and
$n=\sqrt{\epsilon_{\rm r}\mu_{\rm r}}$, respectively. According to
Eqs. (\ref{eq9}) and (\ref{eq16}), the Lorentz transformation for
the optical refractive index is
\begin{equation}
n'=\frac{n+\frac{v}{c}}{1+n\frac{v}{c}}=\tanh\left(n+\frac{v}{c}\right).
\label{eq19}
\end{equation}
Note that here the transformation for the refractive index also
refers only to that of the directions perpendicular to the wave
vector (or ${\bf v}$).
 \\ \\
{\it Fizeau's experiment} The phase velocity of a light in a
moving medium (or seen from a moving frame) is defined as $u'_{\rm
p}=\frac{c}{n'}$. It follows from the expression (\ref{eq19}) that
the phase velocity $u'_{\rm p}$ is
\begin{equation}
u'_{\rm
p}=\frac{c}{n}\left(\frac{1+\frac{nv}{c}}{1+\frac{v}{nc}}\right).
\label{eq20}
\end{equation}
It can be expanded up to the second order in $\frac{v}{c}$, and
the result is
\begin{equation}
u'_{\rm
p}=\frac{c}{n}+\left(1-\frac{1}{n^{2}}\right)v-\frac{c}{n}\frac{v^{2}}{c^{2}}+{\mathcal
O}\left(({v}/{c})^{3}\right).
\end{equation}
It was shown in Fizeau's experiment that the speed of light in a
moving medium is $u'_{\rm
p}=\frac{c}{n}+\left(1-\frac{1}{n^{2}}\right)v$. So, Fizeau's
experimental result is just an effect of the relativistic
transformation of the optical refractive index of the moving
medium.
\\ \\
{\it Superposition principle of phase velocity} The expression
(\ref{eq20}) for the phase velocity $u'_{\rm p}$ can be rewritten
\begin{equation}
u'_{\rm p}=\frac{u_{\rm p}+v}{1+\frac{u_{\rm
p}v}{c^{2}}}=c\tanh\left(\frac{u_{\rm p}}{c}+\frac{v}{c}\right)
\end{equation}
with $u_{\rm p}=\frac{c}{n}$. This, therefore, means that the
phase velocity of light also agrees with the traditional
superposition principle of velocity in SR.
\\ \\
{\it The Lorentz invariance of characteristic impedance of media}
It follows from the expressions (\ref{mu}) and (\ref{epsilon})
that $\sqrt{\frac{\mu'}{\epsilon'}}=\sqrt{\frac{\mu}{\epsilon}}$.
This, therefore, implies that the characteristic impedance of the
medium is a Lorentz scalar, and that the ratio of $E$ to $H$ for a
time-harmonic wave is an invariant.
\\ \\
{\bf 4. Doppler's effect in a moving medium}

Suppose we have a regular anisotropic medium with the rest
refractive index tensor $\hat{n}$, which is moving relative to an
inertial frame K with speed ${\bf v}$ in the arbitrary directions.
The rest refractive index tensor $\hat{n}$ can be written as $
\hat{n}={\rm diag}\left[n_{1}, n_{2}, n_{3}\right] $ with
$n_{i}>0, i=1,2,3$. The wave vector of the electromagnetic wave
under consideration and the relative velocity ${\bf v}$ between
the two frames are chosen to be antiparallel (seen from the
inertial frame K). Thus the wave vector of a propagating wave with
the frequency $\omega$ reads $ {\bf
k}=-\frac{\omega}{c}\left(n_{1}, n_{2}, n_{3}\right)$
 measured by the observer fixed at this medium. In this sense,
one can define a 3-D vector ${\bf n}=\left(n_{1}, n_{2},
n_{3}\right)$, and the wave vector ${\bf k}$ may be rewritten as
${\bf k}=-{\bf n}\frac{\omega}{c}$. Now we analyze the phase
$\omega t-{\bf k}\cdot{\bf x}$ of the above time-harmonic
electromagnetic wave under the following Lorentz transformation
\begin{equation}
{\bf x}'=\gamma\left({\bf x}-{\bf v}t\right), \quad
t'=\gamma\left(t-\frac{{\bf v}\cdot{\bf x}}{c^{2}}\right),
\label{eq1}
\end{equation}
where $\left({\bf x}', t'\right)$ and $\left({\bf x}, t\right)$
respectively denote the spacetime coordinates of the initial frame
K and the system of moving medium, the spatial origins of which
coincide when $t=t'=0$. Here the relativistic factor
$\gamma=\left(1-\frac{v^{2}}{c^{2}}\right)^{-\frac{1}{2}}$. Thus
by using the transformation (\ref{eq1}), the phase $\omega t-{\bf
k}\cdot{\bf x}$ observed inside the moving medium may be rewritten
as the following form by using the spacetime coordinates of K
\begin{equation}
\omega t-{\bf k}\cdot{\bf x}=\gamma\omega\left(1+\frac{{\bf
n}\cdot{\bf v }}{c}\right)t'-\gamma\omega\left(-\frac{{\bf n
}}{c}-\frac{{\bf v}}{c^{2}}\right)\cdot{\bf x}',
\end{equation}
the term on the right-handed side of which is just the expression
for the wave phase in the initial frame K. Hence, the frequency
$\omega'$ and the wave vector ${\bf k}'$ of the observed wave we
measure in the initial frame K are given
\begin{equation}
\omega'=\gamma\omega\left(1+\frac{{\bf n}\cdot{\bf v }}{c}\right),
\quad        {\bf k}'=\gamma\omega\left(-\frac{{\bf n
}}{c}-\frac{{\bf v}}{c^{2}}\right),             \label{eq2eq}
\end{equation}
respectively.

To gain some insight into the meanings of the expression
(\ref{eq2eq}), let us consider the special case of a boost (of the
Lorentz transformation (\ref{eq1})) in the
$\hat{x}_{1}$-direction, in which the medium velocity relative to
K along the positive $\hat{x}_{1}$-direction is $v$. In the
meanwhile, we assume that the wave vector of the electromagnetic
wave is also parallel to the negative $\hat{x}_{1}$-direction. The
modulus of wave vector seen from K is
\begin{equation}
k'=\gamma\omega\left(\frac{n}{c}+\frac{v}{c^{2}}\right).
\end{equation}
If the {\it rest} refractive index of the medium in the
$\hat{x}_{1}$-direction is $n$, its (moving) refractive index in
the same direction measured by the observer fixed at the initial
frame K is of the form
\begin{equation}
n'=\frac{ck'}{\omega'}=\frac{n+\frac{v}{c}}{1+\frac{nv}{c}},
\label{eq3}
\end{equation}
which is a relativistic formula for the addition of ``refractive
indices'' ($\frac{v}{c}$ provides an effective index of
refraction). Obviously, Eq. (\ref{eq3}) is in agreement with Eq.
(\ref{eq19}).

Eq. (\ref{eq2eq}) shows that the mathematical expression for
Doppler's effect in a moving medium is
\begin{equation}
\omega'=\frac{1+\frac{nv}{c}}{\sqrt{1-\frac{v^{2}}{c^{2}}}}\omega.
\label{eqeqDoppler}
\end{equation}
For the vacuum medium ($n=1$), Eq. (\ref{eqeqDoppler}) is reduced
to the one familiar to us, {\it i.e.},
\begin{equation}
\omega'=\sqrt{\frac{c+v}{c-v}}\omega,   \label{longitudinal}
\end{equation}
which is an expression for the longitudinal Doppler's effect.

Recently, a kind of artificial media called left-handed materials
which possess a negative index of refraction attract attention of
many investigators\cite{Pendry1}. Consider a left-handed medium,
the index of refraction of which is $n=-1$. In this case, Eq.
(\ref{eqeqDoppler}) is
$
\omega'=\sqrt{\frac{c-v}{c+v}}\omega $. Compared with
(\ref{longitudinal}), the effect here is shown to be the reversal
of Doppler's shift. Similarly, the reversal of Cerenkov radiation
will also arise in such a negative refractive index material.
\\ \\
{\bf 5. On the quantum vacuum contribution to the momentum of
media}
\\ \\
{\it The linear dispersion relation in a moving electromagnetic
medium} According to the transformation relationships ${\bf
E}'=\gamma \left({\bf E}+{\bf v}\times{\bf B}\right)$ and ${\bf
D}'=\gamma \left({\bf D}+\frac{1}{c^{2}}{\bf v}\times{\bf
H}\right)$, we can obtain $ \epsilon'\left({\bf E}+{\bf
v}\times{\bf B}\right)=\left({\bf D}+\frac{1}{c^{2}}{\bf
v}\times{\bf H}\right)$. Thus we have
\begin{equation}
{\bf D}=\epsilon'{\bf
E}+\left(\frac{\epsilon'\mu-\frac{1}{c^{2}}}{\mu}\right){\bf
v}\times{\bf B}.    \label{D}
\end{equation}
In the same fashion, from ${\bf H}'=\gamma \left({\bf H}-{\bf
v}\times{\bf D}\right)$ and ${\bf B}'=\gamma \left({\bf
B}-\frac{1}{c^{2}}{\bf v}\times{\bf E}\right)$, we have $
\mu'\left({\bf H}-{\bf v}\times{\bf D}\right)={\bf
B}-\frac{1}{c^{2}}{\bf v}\times{\bf E}$, and in consequence obtain
\begin{equation}
{\bf B}=\mu'{\bf H}+\left(\epsilon\mu'-\frac{1}{c^{2}}\right){\bf
E}\times{\bf v}.    \label{B}
\end{equation}
Note that here Eqs. (\ref{D}) and (\ref{B}) are not the same to
the following dispersion relations
\begin{equation}
\left\{
\begin{array}{ll}
&  {\bf D}=\epsilon{\bf
E}+\left(\frac{\epsilon\mu-\frac{1}{c^{2}}}{\mu}\right){\bf
v}\times{\bf B},     \\
&  {\bf B}=\mu{\bf H}+\left(\epsilon\mu-\frac{1}{c^{2}}\right){\bf
E}\times{\bf v}
\end{array}
\right.     \label{feigel}
\end{equation}
derived by Feigel\cite{Feigel}. In Eqs. (\ref{feigel}), the
relativistic transformation is, however, not taken into account
for $\epsilon$ and $\mu$.
\\ \\
{\it Up to the first order in $\frac{v}{c}$ for the permittivity
and permeability of moving medium} In order to treat the quantum
vacuum contribution to the momentum of media, Feigel considered
the first-order approximation (in $\frac{v}{c}$) of the Lagrangian
of the moving electromagnetic system, but did not consider the
first-order (in $\frac{v}{c}$) contribution of $\epsilon$ and
$\mu$. The following expressions
\begin{equation}
\left\{
\begin{array}{ll}
&   \epsilon'_{\rm r}=\epsilon_{\rm r}+\sqrt{\frac{\epsilon_{\rm r}}{\mu_{\rm r}}}
\left(1-\epsilon_{\rm r}\mu_{\rm r}\right)\frac{v}{c}+{\mathcal O}\left(\frac{v^{2}}{c^{2}}\right),     \\
&    \mu'_{\rm r}=\mu_{\rm r}+\sqrt{\frac{\mu_{\rm
r}}{\epsilon_{\rm r}}}\left(1-\epsilon_{\rm r}\mu_{\rm
r}\right)\frac{v}{c}+{\mathcal O}\left(\frac{v^{2}}{c^{2}}\right)
\end{array}
\right. \label{}
\end{equation}
show that the first-order approximation of the optical constants
may also contribute to the momentum of the medium and deserve
consideration.
\\ \\
{\it Lagrangian of moving electromagnetic system} Here, for the
convenient comparison with the result obtained by
Feigel\cite{Feigel}, we adopt the unit system used in Feigel's
paper. So, the relativistic transformations of $\epsilon$ and
$\mu$ take the form
\begin{equation}
\left\{
\begin{array}{ll}
&
\mu'=\sqrt{\frac{\mu}{\epsilon}}\left(\frac{\sqrt{\epsilon\mu}+\frac{v}{c}}{1+\sqrt{\epsilon\mu}\frac{v}{c}}\right),
\\
&
\epsilon'=\sqrt{\frac{\epsilon}{\mu}}\left(\frac{\sqrt{\epsilon\mu}
+\frac{v}{c}}{1+\sqrt{\epsilon\mu}\frac{v}{c}}\right).
\end{array}
\right. \label{}
\end{equation}

In the case of magnetoelectrics, as stated by Feigel, a term
$\frac{1}{\mu}{\bf B}\cdot\hat{\chi}^{\rm T}{\bf E}$ must be added
to the Lagrangian of the moving electromagnetic
system\cite{Feigel}. In a moving magnetoelectric medium, such a
term can be rewritten as
\begin{eqnarray}
& &  \frac{1}{\mu}{\bf B}\cdot\hat{\chi}^{\rm T}{\bf E} \rightarrow    \nonumber \\
&  &  \frac{1}{\mu}{\bf B}\cdot\hat{\chi}^{\rm T}{\bf
E}+\frac{1}{\mu c}\left[{\bf B}\cdot\hat{\chi}^{\rm T}\left({\bf
v}\times {\bf B}\right)+\left({\bf E}\times
{\bf v}\right)\cdot\hat{\chi}^{\rm T}{\bf E}\right]     \nonumber \\
& & +\frac{1}{\mu
c}v\left(\sqrt{\epsilon\mu}-\frac{1}{\sqrt{\epsilon\mu}}\right){\bf
B}\cdot\hat{\chi}^{\rm T}{\bf E},    \label{eqeq37}
\end{eqnarray}
where
\begin{equation}
\hat{\chi}=\left( {\begin{array}{*{20}c}
   {0} & {\chi_{xy}} & {0} \\
   {\chi_{yx}} & {0 } & {0}  \\
   {0} & {0 } & {0}  \\
\end{array}} \right)
\label{eq:scattering}.
\end{equation}
Note that compared with the result derived by Feigel, the
expression $\frac{1}{\mu
c}v\left(\sqrt{\epsilon\mu}-\frac{1}{\sqrt{\epsilon\mu}}\right){\bf
B}\cdot\hat{\chi}^{\rm T}{\bf E}$ in (\ref{eqeq37}) is a new term,
which arises from the relativistic transformation of $\mu$. By
using the relations ${\bf B}\cdot\hat{\chi}^{\rm T}\left({\bf
v}\times {\bf B}\right)=\left({\bf v}\times {\bf
B}\right)\cdot\hat{\chi}{\bf B}={\bf v}\cdot\left[{\bf
B}\times\left(\hat{\chi}{\bf B}\right)\right]$ and $\left({\bf
E}\times {\bf v}\right)\cdot\hat{\chi}^{\rm T}{\bf
E}=\hat{\chi}^{\rm T}{\bf E}\cdot \left({\bf E}\times {\bf
v}\right)=-{\bf v}\cdot\left[{\bf E}\times\left(\hat{\chi}^{\rm
T}{\bf E}\right)\right]$, one can rewrite the term ${\bf
B}\cdot\hat{\chi}^{\rm T}\left({\bf v}\times {\bf
B}\right)+\left({\bf E}\times {\bf v}\right)\cdot\hat{\chi}^{\rm
T}{\bf E}$ in Eq. (\ref{eqeq37}) as ${\bf v}\cdot\left[{\bf
B}\times\left(\hat{\chi}{\bf B}\right)\right]-{\bf
v}\cdot\left[{\bf E}\times\left(\hat{\chi}^{\rm T}{\bf
E}\right)\right]$\cite{Feigel}. Thus, assuming that the velocity
${\bf v}$ of the medium is along the $\hat{{\bf
 z}}$-direction, {\it i.e.}, ${\bf v}=v\hat{{\bf
 z}}$ with $\hat{{\bf
 z}}$ being a unit vector, the Lagrangian of the moving medium is
 thus of the form
\begin{eqnarray}
L_{\rm ME}&=& L_{FM}+\int\frac{{\rm
d}^{3}x}{4\pi}\left(\frac{1}{\mu}{\bf B}\cdot\hat{\chi}^{\rm
T}{\bf E}\right)
    \nonumber   \\
&&  +\frac{1}{\mu c}\int\frac{{\rm d}^{3}x}{4\pi}{\bf
v}\cdot\left\{\left[{\bf B}\times\left(\hat{\chi}{\bf
B}\right)\right]-\left[{\bf E}\times\left(\hat{\chi}^{\rm T}{\bf
E}\right)\right]\right\}
    \nonumber   \\
& & + \frac{1}{\mu c}\int\frac{{\rm d}^{3}x}{4\pi}{\bf
v}\cdot\hat{{\bf
z}}\left(\sqrt{\epsilon\mu}-\frac{1}{\sqrt{\epsilon\mu}}\right){\bf
B}\cdot\hat{\chi}^{\rm T}{\bf E}.   \label{totlag}
\end{eqnarray}
Compared with Eq. (19) in Feigel's paper\cite{Feigel}, the final
expression on the right-handed side of Eq. (\ref{totlag}) in this
Note is a new term. According to the Lagrangian equation (Liquid's
equation) of motion\cite{Feigel}, one can arrive at the following
equation
\begin{eqnarray}
\rho^{0}v\hat{{\bf z}}&=&\frac{1}{4\pi\mu
c}\left[\left({\epsilon\mu-1}\right){\bf E}\times{\bf B}+{\bf
E}\times\left(\hat{\chi}^{\rm T}{\bf E}\right)-{\bf
B}\times\left(\hat{\chi}{\bf B}\right)\right]     \nonumber  \\
& &  -\frac{1}{4\pi\mu
c}\left(\sqrt{\epsilon\mu}-\frac{1}{\sqrt{\epsilon\mu}}\right){\bf
B}\cdot\hat{\chi}^{\rm T}{\bf E}\hat{{\bf z}}.   \label{final}
\end{eqnarray}
Note that the final expression on the right-handed side of Eq.
(\ref{final}) is a new quantum vacuum contribution to the momentum
of media, which has not yet been taken into consideration in
Feigel's work\cite{Feigel}.

\end{document}